\documentclass[journal,transmag]{IEEEtran}

\usepackage{graphicx}
\usepackage{amsmath}
\usepackage{amssymb}
\usepackage{cite}
\usepackage{color}
\usepackage{xcolor}
\usepackage{epstopdf}

\begin{document}

\title{Deep Learning-Based Adaptive Error-Correction Decoding for Spin-Torque Transfer Magnetic Random Access Memory (STT-MRAM)}


\author{\IEEEauthorblockN{Xingwei Zhong\IEEEauthorrefmark{1},
Kui Cai\IEEEauthorrefmark{1},~\IEEEmembership{Senior Member,~IEEE},
Peng Kang\IEEEauthorrefmark{1}, Guanghui Song\IEEEauthorrefmark{2} and Bin Dai\IEEEauthorrefmark{3}}
\IEEEauthorblockA{\IEEEauthorrefmark{1} Science, Mathematics and Technology Cluster, Singapore University of Technology and Design, 487372, Singapore}
\IEEEauthorblockA{\IEEEauthorrefmark{2} State Key Lab of Integrated Services Networks, Xidian University, Xi’an, 710071, China}
\IEEEauthorblockA{\IEEEauthorrefmark{3} School of Internet of Things, Nanjing University of Posts and Telecommunications, Nanjing, 210003, China}
}




\IEEEtitleabstractindextext{
\begin{abstract}
Spin-torque transfer magnetic random access memory (STT-MRAM) is a promising emerging non-volatile memory (NVM) technology with wide applications. However, the data recovery of STT-MRAM is affected by the diversity of channel raw bit error rate (BER) across different dies caused by process variations, as well as the unknown resistance offset due to temperature change. Therefore, it is critical to develop effective decoding algorithms of error correction codes for STT-MRAM. In this paper, we first propose a neural bit-flipping (BF) decoding algorithm, which can share the same trellis representation as the state-of-the-art neural decoding algorithms, such as the neural belief propagation (NBP) and neural offset min-sum (NOMS) algorithm. Hence a neural network (NN) decoder with a uniform architecture but different NN parameters can realize all these neural decoding algorithms. Based on such a unified NN decoder architecture, we further propose a novel deep learning (DL)-based adaptive decoding algorithm whose decoding complexity can be adjusted according to the change of the channel conditions of STT-MRAM. Extensive experimental evaluation results demonstrate that the proposed neural decoders can greatly improve the performance over the standard decoders, with similar decoding latency and energy consumption. Moreover, the DL-based adaptive
decoder can work well over different channel conditions of STT-MRAM irrespective of the unknown resistance offset, with a 50\% reduction of the decoding latency and energy consumption compared to the fixed decoder.
\end{abstract}

}

\maketitle

\IEEEdisplaynontitleabstractindextext

%
\IEEEpeerreviewmaketitle

\section{Introduction}

Owing to its superior features such as the nanosecond read/write speed and low switching energy, spin-torque transfer magnetic random access memory (STT-MRAM) has demonstrated its great potential for various embedded and standalone applications \cite{Chenadvance}. However, the process variation results in a diversity of the raw bit error rate (BER) among different dies of STT-MRAM. In addition, the change of working temperature also causes an unknown offset of the resistance \cite{yuj}. Thus, the existence of the channel raw BER diversity and the unknown channel offset makes the STT-MRAM channel unstable. It also causes difficulty to get the accurate knowledge of the STT-MRAM channel.

In order to mitigate different noises and interference and improve the reliability, error correction codes (ECCs) with short codeword lengths have been applied to STT-MRAM \cite{caij,caic}. For example, a single-error-correcting (71,64) Hamming code is firstly proposed for Everspin's 16 Mb MRAM \cite{everspin}. Bose-Chaudhuri-Hoquenghem (BCH) codes with multiple-error correction capability  are also adopted to further improve the memory storage density of STT-MRAM \cite{caic}.

On the other hand, currently, the deep learning (DL) technique is being actively explored for improving the decoding performance for linear codes \cite{zhongj,nachmanij,carpic}. There are typically three types of standard message-passing based decoding algorithms of linear block codes in the literature, namely, the bit-flipping (BF) decoding algorithm, the min-sum (MS) decoding algorithm, and the BP decoding algorithm, with ascending order of error correction capability and also computational complexity. Both the BP and the MS decoders can be implemented via a trellis graph. By applying the deep unfolding technique, they can be converted into the neural network (NN)-based decoders \cite{zhongj,nachmanij}. For example, a neural normalized-offset reliability-based min-sum (NNORB-MS) decoder is proposed in [5], which can achieve significant performance gain over the standard MS decoder, with a similar decoder structure and time complexity. For the BF decoding algorithm \cite{ryanb}, the reinforcement learning method \cite{carpic} has been adopted to improve its decoding performance. However, being a syndrome based approach \cite{bennatanc}, it cannot be directly implemented by using a trellis structure. Moreover, up till now, all the above NN decoders are used as the fixed decoders to tackle the worst channel condition, and it may result in unnecessary read latency and energy consumption when the channel condition is good and the channel raw BER is low.

In this work, we first propose a neural BF (NBF) algorithm which realizes the BF algorithm \cite{ryanb} using the trellis representation and thereby can share the same structure as the neural belief propagation (NBP) and neural offset MS (NOMS) algorithm \cite{zhongj,nachmanij}. Hence, a NN decoder with a uniform structure but different NN parameters can realize all these neural decoding algorithms. Using such a unified NN decoder architecture, we further propose a novel DL-based adaptive decoding algorithm, whose decoding complexity can be adjusted according to the change of channel conditions of STT-MRAM without the prior knowledge of the channel. We also estimate the decoding latency and energy consumption of the DL-based adaptive decoding algorithm.

The rest of the paper is organized as below. In Section II, we describe the STT-MRAM system and channel model. In Section III, we present the NBF algorithm, and in Section IV, we further propose a DL-based adaptive decoding algorithm.  The experimental evaluation results are illustrated in Section V. Finally, the paper concludes in Section VI.

\section{STT-MRAM System Overview and Channel Model}

\begin{figure}[b]
\centering
\includegraphics[height=0.4\columnwidth]{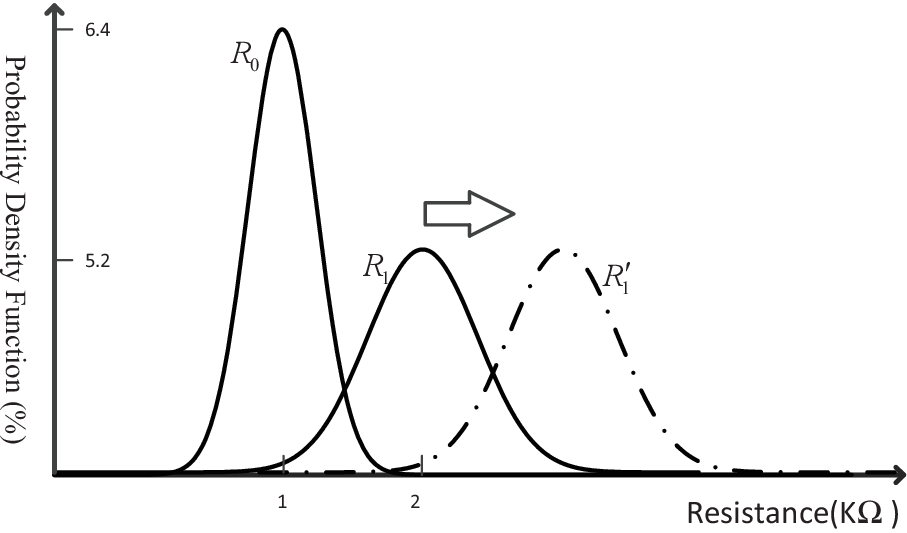}
\caption{The probability density functions (PDFs) of resistances for STT-MRAM: $R_{1}$ represents the initial high resistance, and $R_{1}^{'}$ is the shifted high resistance caused by the drop of temperature, and $R_{0}$ represents the fixed low resistance.}
\label{pdf_model}
\end{figure}

Process imperfection is a main factor that affects the data reliability of STT-MRAM. { In particular, the high and low resistances of the magnetic tunneling junction (MTJ) caused by the quantum mechanical tunneling are inversely related to its cross-section area, and they change exponentially with the tunneling oxide thickness \cite{caic}. Therefore, the process imperfection will result in widened distributions of the high and low resistance states and their overlapping \cite{caij,caic}. Meanwhile, it has also been found that temperature variation will affect the resistance distributions of STT-MRAM \cite{wuj}. According to \cite{wuj}, with the increase of temperature, the high resistance $R_{1}$ increases while the low resistance $R_{0}$ hardly changes, leading to a reduction of the tunneling magnetoresistance
(TMR) ratio defined as ${\rm TMR}=(R_{1}-R_{0})/R_{0}\times 100\%$. Experimental measurement results indicate that the drop of TMR with the increase of temperature is almost linear \cite{wuj}.} {  However, in practical situations it is difficult to predict online instantaneously the change of TMR over different temperatures accurately, thereby the TMR change is unknown to the memory sensing circuit and the ECC decoder.
Therefore, it will lead to more memory read errors if the threshold resistance ({\it i.e.} memory sensing reference current/voltage) that differentiates the two resistance states is still kept the same, and if the conventional ECC decoder is adopted. Such an effect is illustrated by Fig. \ref{pdf_model}.}

By following \cite{zhongj,meic} and according to stochastic characteristics of the low and high resistance states, the STT-MRAM channel can be modelled as

\begin{equation} \label{model}
y_{i}=r_{i}+n_{i}+b_{i},
\end{equation}
where $y_{i}$ denotes the read back resistance value of the $i$-th memory cell, with $i=1,\cdots,N$, and $N$ represents the length of the linear block code. The nominal resistance value $r_{i}$ is determined by the stored input bit $x_i \in \{0,1\}$. Thus, $r_{i}=\mu_0$ for $x_i=0$, and $r_{i}=\mu_1$ for $x_i=1$. We also include an independent and identically distributed (i.i.d) Gaussian variable $n_{i} \in \mathbb{R}$ with zero mean and variance $\sigma^2$ to model the resistance variation of STT-MRAM cell caused by process imperfection. Note that $n_{i}$ can be also non-Gaussian distributed. {  In this work, we use $b_i$ to denote the unknown resistance offset caused by the change of temperature. Since only $R_{1}$ will be affected by the unknown offset, we let $b_i=0$ for $x_i=0$. According to the fact that temperature will affect the resistance of each memory cell randomly, we assume that the offset $b_i$ for $x_i=1$ is an i.i.d Gaussian variable $\mathcal{N}(\mu_{b}, \sigma^{2}_{b})$  with mean of $\mu_{b}$ and standard deviation of $\sigma_b$. Thus, the values of $\mu_{b}$ and $\sigma_b$ can be changed to reflect the change of TMR with different temperatures in different environments.}

{  We remark that the non-linear or data-dependent distortions of STT-MRAM can also be included in the channel model. We will leave it as our future work. Nevertheless, due to the lack of sufficiently large amount of measured data of STT-MRAM, the channel model of (1) is only used to generate the training and testing data sets to train and test the DL-based decoders. Our proposed DL-based adaptive decoding algorithm does not need to know this channel model.}

\section{Neural Bit-flipping (BF) Algorithm}

\begin{figure}[b]
\centering
\includegraphics[height=0.7\columnwidth]{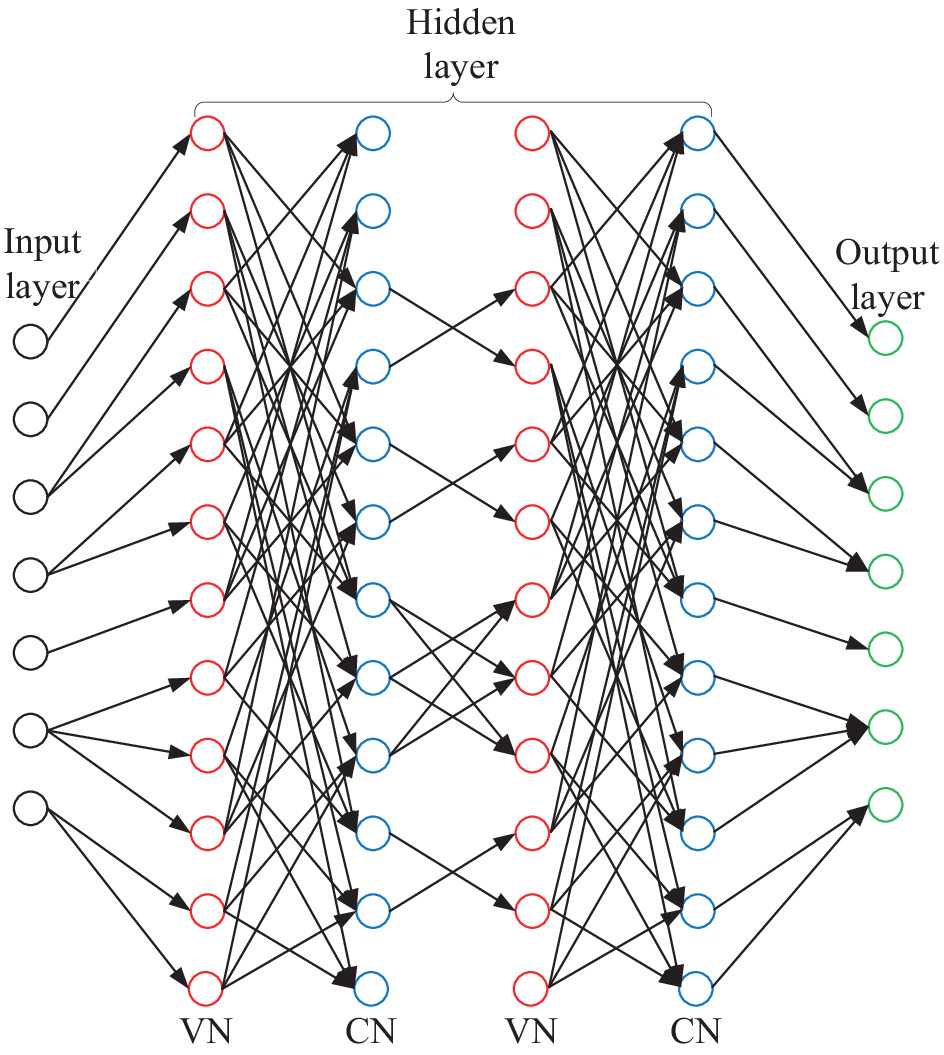}
\caption{DNN architecture for the (7,4) Hamming code with 4 hidden layers corresponding to 2 full decoding iterations.}
\label{DNN_architecture}
\end{figure}

The hard-decision decoding algorithms, such as the BF decoding algorithm, can further lower the decoding complexity and latency compared to the soft-decision decoding algorithms such as the BP and MS algorithms. In order to derive the NBF algorithm which shares the same structure as the NBP and NOMS algorithms \cite{zhongj,nachmanij}, we first present the trellis representation of the BF algorithm as below.

(a) For odd index $t$ corresponds to the variable node (VN) layer:
\begin{equation} \label{bf_vn_update}
\lambda_{k}^{(t)} = \sum_{c \in M(k)} \hat{u}_c H_{kc}\ ({\rm mod}~2),
\end{equation}
where $\lambda_{k}^{(t)}$ represents the $k$-th variable-based message at the $t$-th layer, $M(k)$ is the set of CNs connected to VNs, $\hat{u}_c$ is the $c$-th hard decision of the channel output, and $H_{kc}$ is the parity-check matrix of the code.

(b) For even $t$ corresponds to the check node (CN) layer:
\begin{equation} \label{bf_cn_update}
\varepsilon_{c}^{(t)} = -\sum_{k \in N(c)} (1-2\lambda_{k}^{(t)}),
\end{equation}
where $\varepsilon_{c}^{(t)}$ is the $c$-th check-based message at the $t$-th layer, and  $N(c)$ is the set of VNs connected to CNs. The bit $\hat{u}_c$ for $c=\arg\max_{1\leq c \leq N} \varepsilon_{c}^{(t)}$ will be flipped.

Based on the trellis presentation of the BF algorithm described by (\ref{bf_vn_update}) and (\ref{bf_cn_update}), we propose the NBF decoder by adding the weights to the edges. For the proposed NBF decoder, (\ref{bf_vn_update}) is unchanged while (\ref{bf_cn_update}) is replaced by:

\begin{equation} \label{nbf_cn_update}
\varepsilon_{c}^{(t)} = -\sum_{k \in N(c)} (1-2\lambda_{k}^{(t)})w_{k,c}^t,
\end{equation}
where $w_{k,c}^t$ is the learnable weight for edge ($k,c$) in the $t$-th layer. The weight can then be optimized by the NN.

The above proposed NBF algorithm can share the same structure as the NBP and NOMS algorithm and hence the corresponding NN decoder with a uniform structure but
different NN parameters can realize all these neural decoding algorithms. As an example, the deep neural network (DNN) architecture for the (7,4) Hamming code with 4 hidden layers corresponding to 2 full decoding iterations is illustrated by Fig. \ref{DNN_architecture}. It can be shared by the NBF, NOMS, and NBP decoders.

\section{DL-Based Adaptive Decoding Algorithm}
Based on the unified NN decoder architecture presented in the previous section, we further propose a novel DL-based adaptive decoding
algorithm in this section, whose decoding complexity can be adjusted according to the change of the channel conditions of STT-MRAM
without the prior knowledge of the channel.

\begin{figure}[h]
\centering
\includegraphics[width=3.3in, height=4.1in]{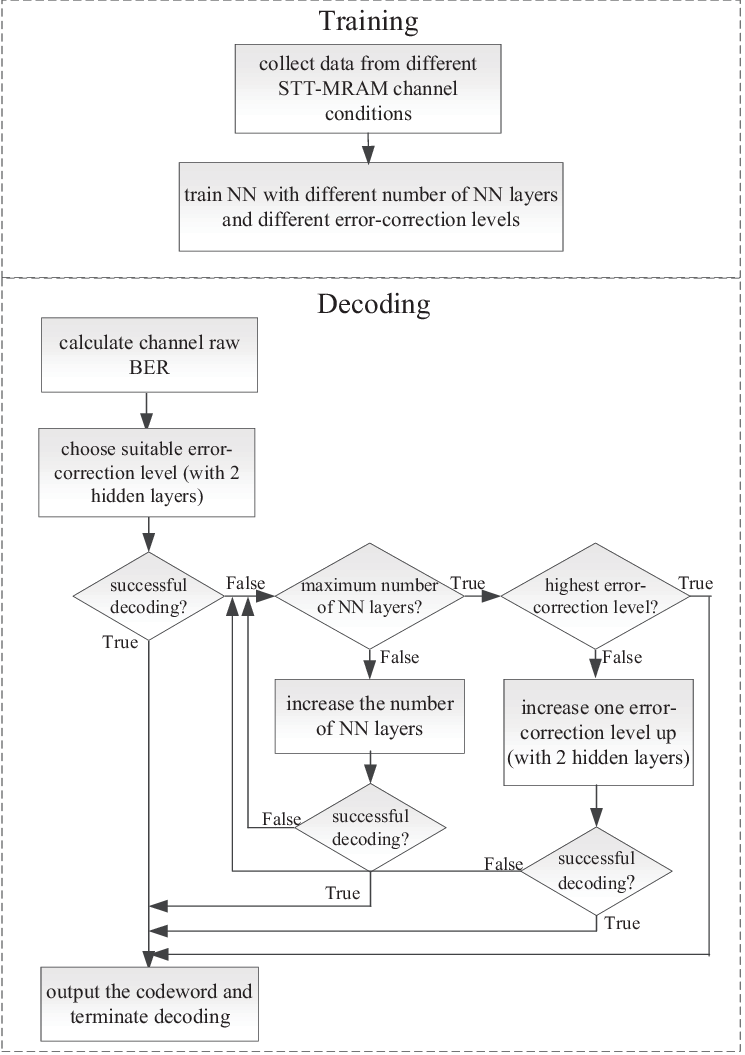}
\caption{Block diagram of the proposed DL-based adaptive decoding algorithm.}
\label{block_graph_algorithm}
\end{figure}

The key idea is as follows. We use several types of decoders with the trellis representation, such as the BF decoder, the MS decoder, and the BP decoder, for different levels of error correction. By using the deep unfolding technique, all these decoders can be converted into a NN decoder with a uniform structure but different NN parameters, with the simpler decoders having some NN parameters to be set to zeros \cite{zhongj}. In order to reduce the energy consumption and read latency, we only invoke stronger levels of error-correction or adopt more NN layers when necessary. That is, when the channel condition is good and the channel raw BER obtained from the channel detector (e.g. soft-output channel detector described in \cite{zhongj}) is low, we use simpler decoders that have fewer NN parameters and adopt less number of NN layers; when the channel condition is getting worse and the channel raw BER is higher, we invoke higher level of error-correction and apply more complicated decoders that have more NN parameters, and also adopt more number of NN layers for decoding. A block diagram of the proposed DL-based adaptive decoding algorithm is illustrated in Fig. \ref{block_graph_algorithm}. The general process of the DL-based adaptive decoding algorithm is summarized as follows.

\pmb{Step 1}: Set the NBF, NOMS, and NBP as the error-correction level 1, error-correction level 2, and error-correction level 3, respectively. Train these NN decoders for different levels of error-correction with different number of NN layers using the data collected from different channel conditions of STT-MRAM.

\pmb{Step 2}: During decoding, use the raw BER of the STT-MRAM channel as an indicator to choose a specific error-correction level and the corresponding trained NN decoder. Gradually increase the number of NN layers if the decoding is not successful. {   Note that the channel raw BER can be obtained from the memory sensing circuit ({\it i.e.} channel detector) for a small amount of cells where the information stored in is known, such as the reference cells.}

\pmb{Step 3}: If the decoding fails in \pmb{Step 2} with the maximum number of NN layers, increase one level up for error-correction, and use the trained NN decoder with better error-correction performance and higher complexity to do decoding. Gradually increase the number of NN layers if the decoding is not successful. Otherwise, output the decoded codeword and terminate the decoding process.

\pmb{Step 4}: Repeat \pmb{Step 3} if the decoding is not successful until the decoding reaches the highest error-correction level with the maximum number of NN layers. Output the decoded codeword and terminate the decoding process.

{   The computational complexity of the three types of neural decoders (NBF, NOMS, NBP) can be analyzed in terms of the numbers of additions, comparisons, multiplications, XORs, and Tanh operations involved in the decoding algorithms [7]. The results are summarized in Table \ref{Table 7.1}. In the table, $E$ denotes the number of edges, $C$ is the number of CNs, $V$ is the number of VNs, and $I$ represents the number of decoding iterations. }

\begin{table}[h]
\small
\centering
\caption{{ Computational complexity comparison of various decoders.}}
\label{Table 7.1}
 
\begin{tabular}{l|l|l|l}

Operation              & NBF & NOMS & NBP \\ \hline
Addition       & $\approx 2EI$  & $3EI$ & $2EI$   \\ \hline
Comparison       & $(V-1)I$  & $\approx 2EI$ & 0   \\ \hline
Multiplication      &  $CI$  & 0 & $\approx 3EI$   \\ \hline
XOR        & $I$  & $\approx 2EI$ & 0   \\ \hline
Tanh      &  0 & 0 & $2EI$   \\ \hline

\end{tabular}
\end{table}

\section{Performance Evaluations}

\subsection{Latency and Energy Consumption Evaluations}
In this sub-section, we evaluate the latency and energy consumption of various decoders. Instead of using hardware such as the field-programmable gate array (FPGA), digital signal processors (DSP), or the general purpose processors
(GPPs), we follow \cite{leec} and adopt a fast and flexible simulator named Wattch \cite{brooksj} for the measurement. Wattch is a framework for analyzing microprocessor energy consumption and latency at the architecture-level. Compared with the existing layout-level energy tools, Wattch can achieve around 1000 times faster and maintain above 90$\%$ accuracy \cite{brooksj}. To reflect the common processor architectures, the parameters of the Wattch are set with 600 MHz clock rate, 4 issue width, 16 window size, 32 virtual registers, 16 physical registers, and 64 datapath \cite{leec}. During evaluations,  different decoders with the same iteration number $I=5$ are fed into the Wattch, which follows the BER evaluations presented in the next sub-section.

\begin{figure}[t]
\centering
\includegraphics[height=0.55\columnwidth]{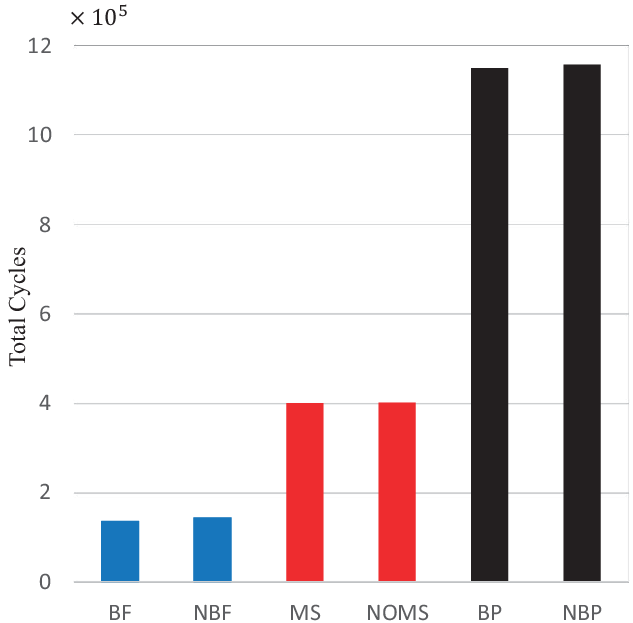}
\caption{Total cycles of different decoders with $I=5$.}
\label{total_cycle}
\end{figure}

\begin{figure}[b]
\centering
\includegraphics[height=0.55\columnwidth]{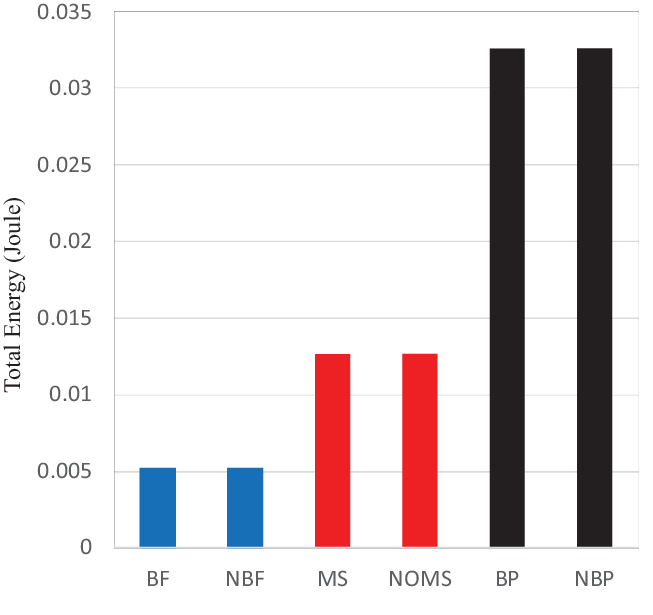}
\caption{Energy consumption of different decoders with $I=5$.}
\label{total_energy}
\end{figure}

Fig. \ref{total_cycle} shows the total number of processor cycles ({\it i.e.} an indication of decoding latency) and Fig. \ref{total_energy} illustrates the total energy consumption of different decoders with $I=5$, which are reported by Wattch. It can be observed that the neural decoders (NBP, NOMS, NBF) lead to almost the same number of cycles and energy consumption as their standard counterparts (BP, MS, BF) with the same iteration number. Moreover, the latency (total number of cycles) of NBF is about 1/3 of NOMS and 1/8 of NBP. Thus, we have $T_{NBF}=1/3T_{NOMS}=1/8T_{NBP}$. For the energy consumption, the NBF is about 1/2 of NOMS and 1/6 of NBP, and hence $E_{NBF}=1/2E_{NOMS}=1/6E_{NBP}$.

\begin{figure}[b]
\centering
\includegraphics[height=0.7\columnwidth]{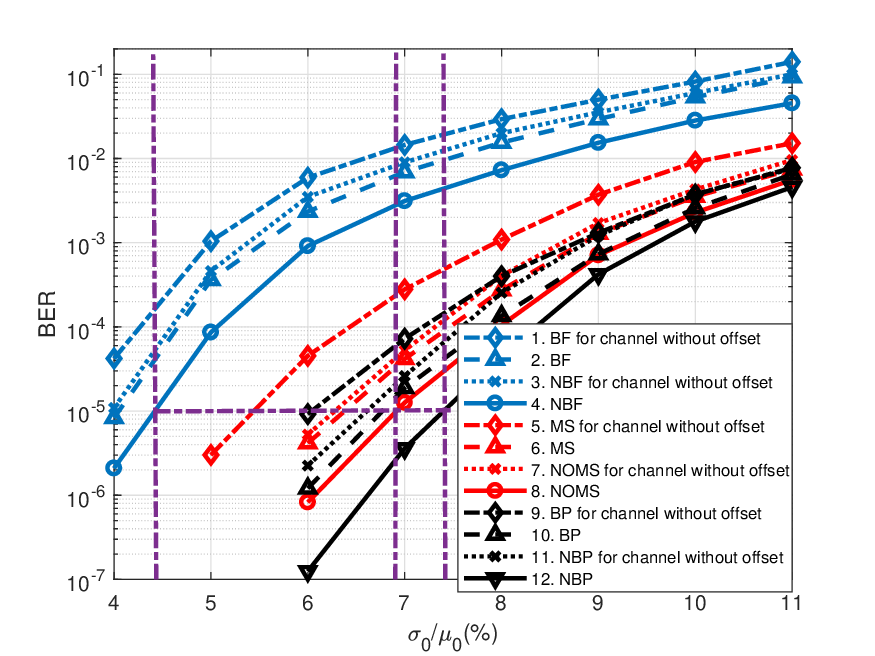}
\caption{BERs of different decoders for the channel with resistance offset of $\mu_{b} = -0.2 k\Omega$ and  $\sigma_{b} / \mu_{1}  = 3 \%$.}
\label{simulation_2}
\end{figure}

\subsection{BER Evaluations}

In this sub-section, we use computer simulations to evaluate the BERs of various decoders over different resistance spreads and resistance offsets. In our experiment, the various neural decoders are implemented through the Tensorflow framework to optimize the learnable weights. The details of the NN settings are given in Table I. The all-zero codeword is adopted for training and the random codeword is adopted for testing of the NN. In the simulations, we follow \cite{zhongj} and adopt the (71,64) Hamming code as the ECC. We adopt the nominal values of the high and low resistance states of STT-MRAM of $\mu_{0} = 1k\Omega$ and $\mu_{1} = 2k\Omega$, respectively. Moreover, we assume that the mean normalized resistance spreads $\sigma_{0} / \mu_{0}=\sigma_{1} / \mu_{1}$ according to the features of the memory fabrication process. We use $\sigma_{0} / \mu_{0}$ as an indication of different process vatiations.

\begin{table}[h]
\small
\caption{Network settings for the proposed DL-based adaptive decoders.}
\label{Table 7.2}
\scalebox{0.95}{
\begin{tabular}{l l l l}
\hline \hline
Decoder Type       & NBF                  & NOMS           & NBP            \\ \hline
Training Samples   & $4\times 10^4 N$ & $4\times 10^4 N$ &                         $4\times 10^4 N$ \\ \hline
Testing Samples    & $1\times 10^6 N$& $1\times 10^6 N$ & $1\times 10^6 N$                            \\ \hline
Mini-batch Size    & 120 & 120 & 120                               \\ \hline
Learning Rate      & 0.000003             & 0.1            & 0.05           \\ \hline
Optimizer          & \multicolumn{3}{l}{Adam with multi-loss optimization \cite{nachmanij}} \\ \hline \hline
\end{tabular}
}
\end{table}

For the case where the channel has an unknown offset with mean of $\mu_{b} = -0.2k\Omega$ and normalized spread of $\sigma_{b}/\mu_{1}=3\%$ in Fig. \ref{simulation_2} illustrates the BER performance of various standard and neural decoders over different resistance spreads, where the channel has an unknown resistance offset with mean of $\mu_{b} = -0.2k\Omega$ and normalized spread of $\sigma_{b}/\mu_{1}=3\%$. From Fig. \ref{simulation_2}, we observe that the BERs of the standard BF, MS, and BP decoders (Curves 1,5,9) are poor, when the decoding is carried out using the LLRs without taking into account the resistance offset (\textbf{the decoders only know the original channel without offset}). However, by using the NN-based soft information generation method in \cite{zhongj}, accurate channel soft information can be obtained and sent to the decoders, and the performance improvement can be seen from Curves 2,6,10, for the BF, MS, and BP decoders, respectively. {Moreover, when trained by the LLRs without taking into account the resistance offset, the neural decoders's performance (Curves 3,7,11) is worse than that of the corresponding standard decoders (Curves 2,6,10).} However, when trained by the accurate channel soft information, all the neural decoders (Curves 4,8,12) achieve obvious BER gain over the corresponding standard decoders (Curves 2,6,10). Among the neural decoders, the NBP (Curve 12) achieves the best BERs with the highest latency and power consumption as shown in Section V.A, while the NBF (Curve 4) provides the worst BERs, but with the lowest latency and power consumption.



\subsection{Latency and Energy Consumption Reduction}

With an assumed target BER of $\rm {BER}=1\times 10^{-5}$, it can be seen that instead of using a fixed NBP decoder for all the cases when the mean normalized resistance spreads $\sigma_{0} / \mu_{0}$ is less than 7.45\%, we can use the lower level neural decoders for cases with smaller resistance spread (we can obtain the resistance spread level from the channel raw BER as stated in Step 2 of our agorithm). That is, use the NBF decoder for $\sigma_{0} / \mu_{0}<4.45\%$, the NOMS decoder for $4.45\%<\sigma_{0} / \mu_{0}<6.85\%$, and only invoke the NBP decoder for $6.85\%<\sigma_{0} / \mu_{0}<7.45\%$. In this way, we can minimize the probability that the neural decoders with high computational complexity is invoked. Considering the three cases of $\sigma_{0} / \mu_{0}=4.45\%$, $\sigma_{0} / \mu_{0}=6.85\%$, and $\sigma_{0} / \mu_{0}=7.45\%$, if using a fixed NBP decoder, the corresponding decoding latency is $3T_{NBP}$ which equals $24T_{NBF}$ according to the experimental evaluations of Section V.A. However, if adopting the adaptive decoding, the average decoding latency is $T_{NBF}+T_{NOMS}+T_{NBP}=12T_{NBF}$. Therefore, the adaptive decoding achieves a 50\% reduction of the latency. Similarly, the adaptive decoding consumes an average energy of $E_{NBF}+E_{NOMS}+E_{NBP}=9E_{NBF}$, while the fixed NBP decoder consumes the energy of $3E_{NBP}=18E_{NBF}$. Hence the energy consumption is also reduced by half. Furthermore, all the three neural decoders are realized in a single decoder structure, and hence the silicon area is also minimized.

\section{Conclusions}
In this work, we have investigated DL-based decoding algorithms of ECCs to tackle the process variation induced resistance distributions, as well as the the variation of temperature induced unknown resistance offset of STT-MRAM. We have first proposed a NBF algorithm which can share the same trellis representation as the NBP and NOMS algorithms. Hence, a NN decoder with a uniform structure but different NN parameters can realize all the three neural decoding algorithms. We have further proposed a DL-based adaptive decoding algorithm which can choose the appropriate neural decoder and the NN layers for a specific channel condition, so as to minimize the read latency and energy consumption. We have carried out extensive experimental evaluations, and the results demonstrate that the proposed neural
decoders can effectively improve the performance over the standard decoders, with similar read latency
and energy consumption. Moreover, the DL-based adaptive
decoder can work well over different channel conditions of
STT-MRAM irrespective of the unknown resistance offset, with a 50\% reduction of the decoding latency and energy
consumption compared to the fixed decoder. Therefore, we conclude that our proposed DL-based adaptive
decoding algorithms can effectively mitigate the impairments of STT-MRAM and support its applications with low latency and energy consumption requirements.

\section*{Acknowledgement}
This work is supported by Singapore MOE Tier 2 fund T2EP50221-0036 and RIE2020 Advanced Manufacturing and Engineering (AME) programmatic grant A18A6b0057.


%
%
%
%
%

\end{document}